\documentclass[9pt,twocolumn,twoside]{osajnl}

\pdfoutput=1
\usepackage{subcaption,graphicx,trimclip}
\usepackage[nameinlink]{cleveref}

\journal{ol}
\setboolean{shortarticle}{true}
\usepackage[utf8]{inputenc}
\usepackage[english]{babel}
\usepackage{dcolumn}
\usepackage{bm}
\usepackage{hyperref}
\usepackage{graphicx}
\usepackage{textcomp}
\usepackage{lipsum}
\usepackage{pdfcomment}
\usepackage{siunitx}
\usepackage{bold-extra}
\usepackage{ulem}

\let\titleoriginal\title           % save original \title macro
\renewcommand{\title}[1]{          % substitute for a new \title
    \titleoriginal{#1}%               % define the real title
    \newcommand{\mytitle}{#1}        % define \thetitle
}

\newcommand{\greg}[1]{#1}

\title{Impact of Stoichiometric Silicon Nitride Growth Conditions on Dispersion Engineering of Broadband Microresonator Frequency Combs}

\author[1,2*]{Gregory Moille}
\author[2]{Daron Westly}
\author[2]{Gregory Simelgor}
\author[1,2]{Kartik Srinivasan}
\affil[1]{Joint Quantum Institute, NIST/University of Maryland, College Park, USA}
\affil[2]{Microsystems and Nanotechnology Division, National Institute of Standards and Technology, Gaithersburg, USA}
\affil[*]{Corresponding author: gmoille@umd.edu}

\begin{abstract}
Microresonator frequency combs, or microcombs, have gained wide appeal for their rich nonlinear physics and wide range of applications. Stoichiometric silicon nitride films grown via low-pressure chemical vapor deposition (LPCVD), in particular, are widely used in chip-integrated Kerr microcombs. Critical to such devices is the ability to control the microresonator dispersion, which has contributions from both material refractive index dispersion and geometric confinement. Here, we show that modifications to the LPCVD growth conditions, specifically the ratio of the gaseous precursors, has a significant impact on material dispersion and hence the overall microresonator dispersion. In contrast to the many efforts focused on comparison between Si-rich films and stoichiometric (Si$_3$N$_4$) films, here we entirely focus on films that are within the nominally stoichiometric growth regime. We further show that microresonator geometric dispersion can be tuned to compensate for changes in the material dispersion.    
\end{abstract}

\dates{\today}
\setboolean{displaycopyright}{true}

\begin{document}
\maketitle
\thispagestyle{fancy}

\begin{figure}[!t]
    \begin{center}
        \includegraphics{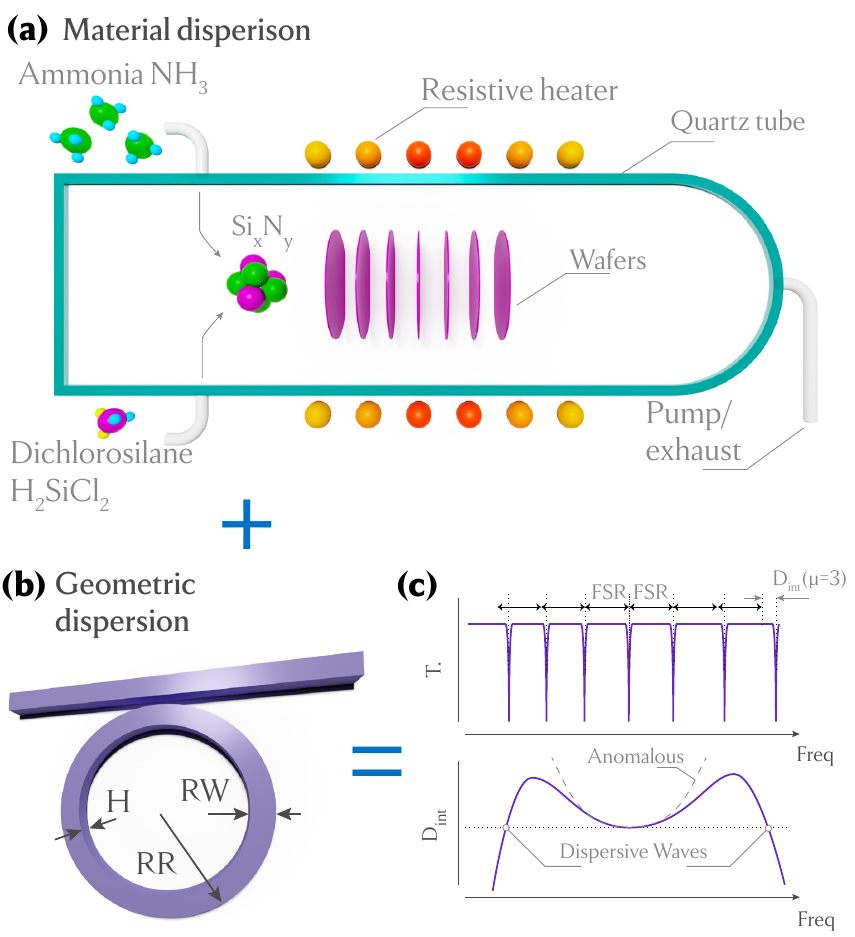}
        \caption{\label{fig:1}\textbf{(a)} LPCVD system, where SiO\textsubscript{2} on Si wafers are loaded within a quartz furnace tube which is heated to a temperature of 775\si{\degree}C in our case. The two precursor gases, are injected into the chamber at controllable flow levels, allowing change of the gas ratio in the reaction and impacting the material refractive index and its dispersion. \textbf{(b)} Microring resonator, which is coupled to a waveguide for injection of pump light and collection of generated comb light. The three geometric parameters, ring radius $RR$, thickness $H$, and ring width $RW$ determine the geometric dispersion contribution. \textbf{(c)} Both material and geometric dispersion contribute to the total dispersion of the resonato, allowing anomalous dispersion regime - free spectral range increasing with the frequency (upper panel) - preferable for microcomb operation. In addition, this dispersion determines the spectral position of any dispersive waves, which appear at zero-crossings of the integrated dispersion $D_\text{int}$ (lower panel).}
    \end{center}
\end{figure}

The ability to realize frequency comb generation in integrated photonic platforms through dissipative Kerr soliton (DKS) formation in $\chi^{(3)}$ microresonators has opened up numerous applications in timekeeping, communications, and spectroscopy~\cite{diddamsOpticalFrequencyCombs2020}. Recent demonstrations combining DKS microcombs with chip-scale lasers~\cite{sternBatteryoperatedIntegratedFrequency2018}, to the extent of realizing octave-spanning bandwidth~\cite{brilesHybridInPSiN2021a}, highlight the potential for field deployment of such systems. This is particularly true for silicon nitride (SiN), which not only has been utilized in numerous demonstrations~\cite{gaeta_photonic-chip-based_2019}, but has been shown to fit within foundry-like fabrication process flows suitable for mass production~\cite{liuHighyieldWaferscaleFabrication2021}. Successful use of SiN within applications hinges on its large Kerr nonlinearity, broadband optical transparency, and reproducible and controllable dispersion, the latter \greg{is} of which is particularly critical for broadband applications such as the generation of octave-spanning combs~\cite{liOctavespanningMicrocavityKerr2015,pfeifferOctavespanningDissipativeKerr2017,moilleUltraBroadbandSolitonMicrocomb2021}. The microcomb bandwidth and shape for a given material is entirely defined through the resonator dispersion, which can be decomposed in two elements: the material dispersion and the geometric dispersion (\cref{fig:1}), the former defined during the material growth, and the latter relying on accurate dimensions of the resonator compared to the theoretical design. SiN growth for microcombs is typically done via low pressure chemical vapor deposition (LPCVD) using two precursors gases, ammonia (NH\textsubscript{3}) and dichlorosilane (DCS or SiH\textsubscript{2}Cl\textsubscript{2}), where the ratio between these two gases defines the composition of the deposited film. Most Kerr comb \greg{works have} focused on achieving stoichiometric films, i.e., Si$_3$N$_4$, though a substantial amount of work has been conducted in studying silicon rich films~\cite{dizajiSiliconrichNitrideWaveguides2017,kruckelLinearNonlinearCharacterization2015,tanNonlinearOpticsSiliconrich2018,yeLowlossHighQSiliconrich2019}, in part due to the higher nonlinear coefficient in this regime, with its refractive index dispersion also having been thoroughly studied~\cite{kruckelLinearNonlinearCharacterization2015}. Yet, stoichiometric films are often preferred due to linear losses that are typically lower and a wider optical transparency window, the latter of particular importance considering the possibility of nonlinear absorption. Interestingly, the gas ratio between NH$_3$ and DCS at which the \greg{stochiometric} condition is reached remains unclear; often NH$_3$:DCS $>2:1$ is cited~\cite{diraniAnnealingfreeSi3N4Frequency2018a}, though larger NH$_3$:DCS ratios have also been considered~\cite{lukeBroadbandMidinfraredFrequency2015,xuanHighQSiliconNitride2016}. Here, we study LPCVD-deposited SiN films within this nominally stoichiometric regime with a NH$_3$:DCS gas ratio $>2:1$. We find that the refractive index dispersion continues to change as the NH$_3$:DCS ratio is increased, with 3:1, 5:1, 7:1, and 15:1 ratios considered. We consider the impact of this varying material dispersion on the integrated dispersion of microresonators fabricated in these LPCVD-deposited films, and show through experimental generation of DKS microcomb states that differences in material dispersion between the films can be somewhat compensated through geometric dispersion, in particular, the thickness of the SiN film.
 
\begin{figure}[!t]
    \begin{center}
        \includegraphics{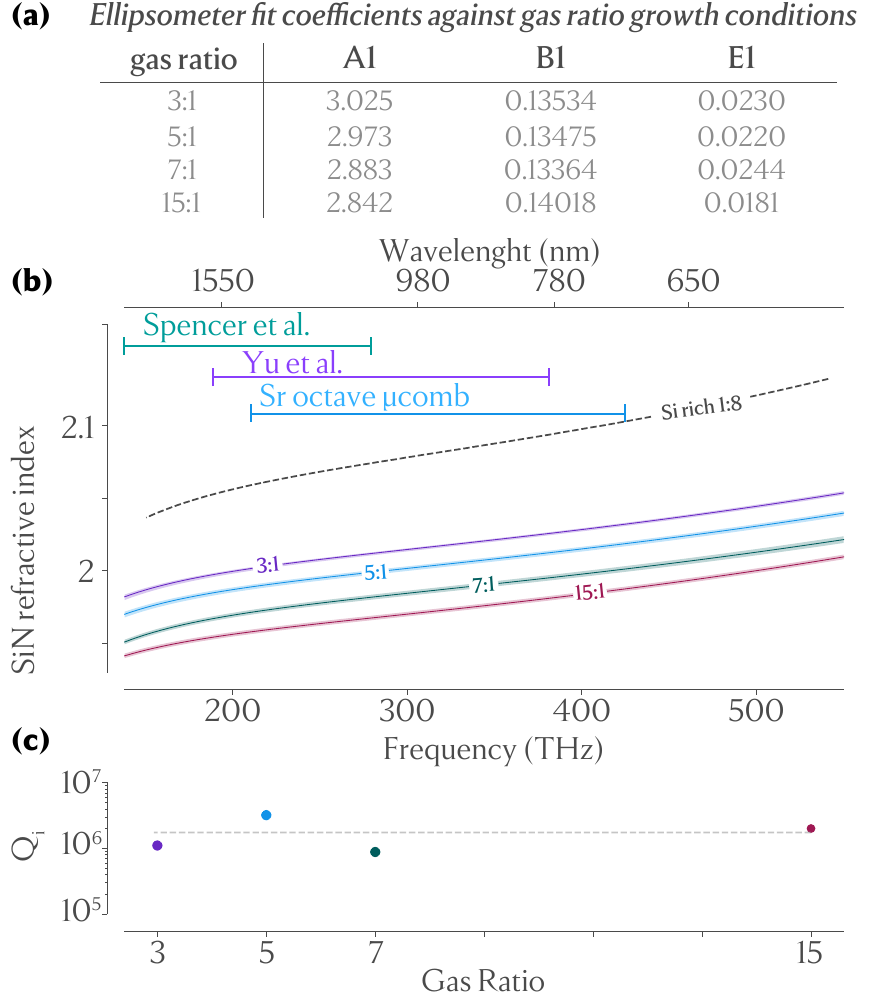}
        \caption{\label{fig:2}\textbf{(a)} Ellipsometer fit coefficients for the different SiN growth conditions (varying gas ratio), where the films are deposited on silicon substrates and the ellipsometer data is fit to a standard Sellmeier model. \textbf{(b)} Frequency-dependent refractive index obtained from the previous fits for gas ratios of 3:1 (purple), 5:1 (cyan), 7:1 (teal) and 15:1 (red). \greg{The transparent areas account for the uncertainty of about 1\% in the ellipsometer fitting coefficients}. The Si-rich refractive index data (grey dashed) has been extracted from~ref.~\cite{xuanHighQSiliconNitride2016}. The octave spanning frequency ranges  for different microcombs are indicated on the top left, and correspond to experimental data of refs\cite{spencerOpticalfrequencySynthesizerUsing2018,yuTuningKerrsolitonFrequency2019,moillePostProcessingDispersionEngineering2020a}. \textbf{(c)} Average extracted intrinsic quality factor of ring resonators with the same geometry but created under different growth conditions. \greg{The standard variation is smaller than the displayed data point}}
    \end{center}
\end{figure}
 \begin{figure*}[t]
    \begin{center}
        \includegraphics{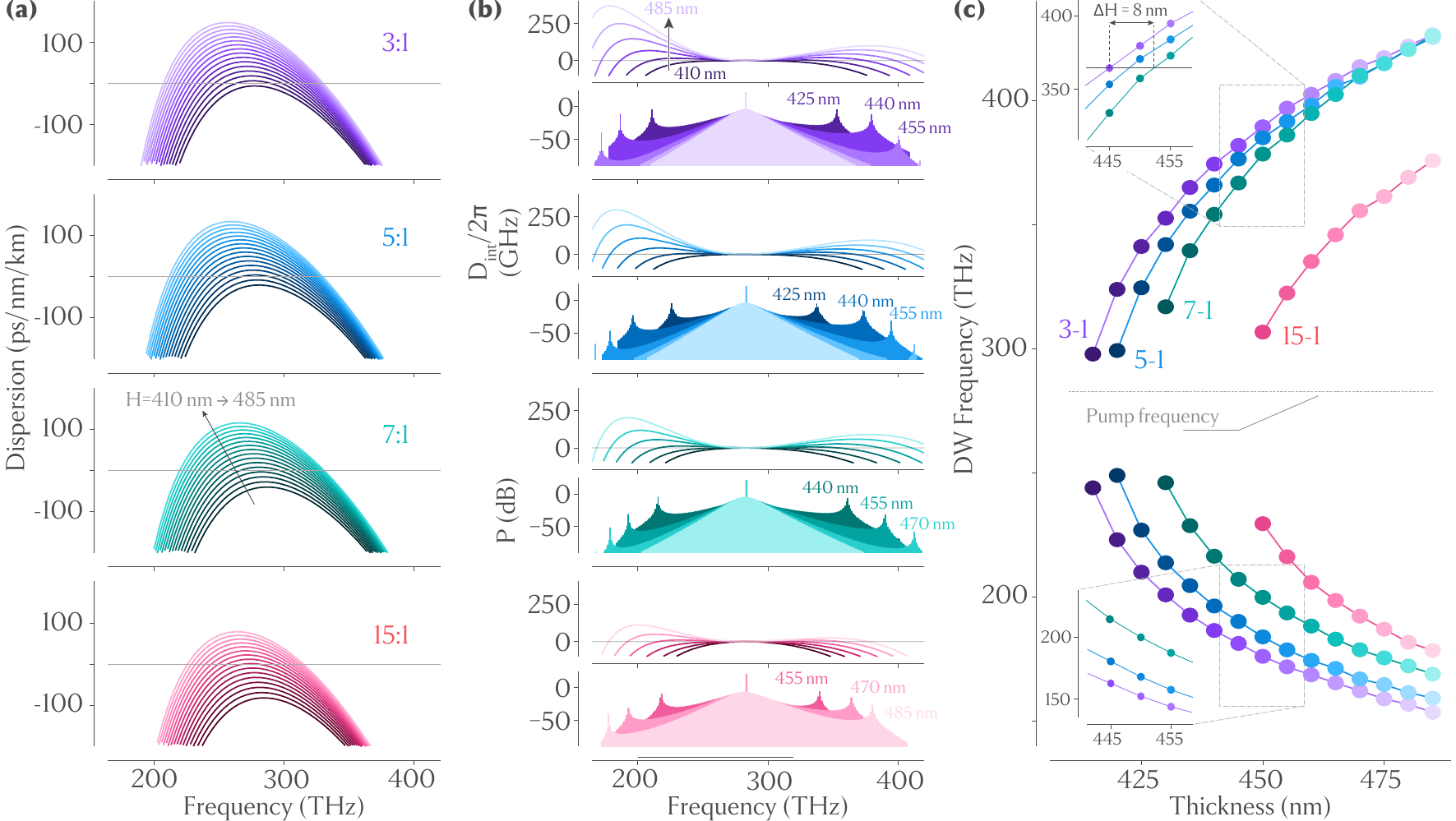}
        \caption{\label{fig:3} \textbf{(a)} \greg{Numerical simulations of the} dispersion parameter for a microring resonator with $RR=23$~{\textmu}m and $RW=1060$~nm for a thickness that varies from 410~nm (darker) to 485~nm (lighter) considering different gas ratio, 3:1 (purple), 5:1 (cyan), 7:1 (teal) and 15:1 (red). \textbf{(b)} Integrated dispersion for the same set of geometries and gas ratios (top panel), with their corresponding LLE simulated spectra (bottom panel). \textbf{(c)} Position of the DWs, obtained from the zero-crossings of the integrated dispersion, as a function of thickness for the different gas ratios of interest, considering a pump frequency at 281~THz. The insets highlight the possibility to compensate for the DW shift due to the material dispersion by adjusting the geometric dispersion.}
    \end{center}
\end{figure*}
% --- Ellipsometry and Q ---- -
  Growth of SiN is conducted through LPCVD onto a 3~{\textmu}m thick silicon dioxide (SiO\textsubscript{2}) layer that has been grown via thermal oxidation of a 100~mm silicon wafer. The precursor gases, as stated previously, are NH\textsubscript{3} and DCS, where the NH\textsubscript{3}/DCS ratio is adjusted with a fixed DCS flow of \greg{$8.3\times10^{-7}$~m\textsuperscript{3}.s\textsuperscript{-1} (50~sccm)}. We use a fixed temperature of 775~\si{\degree}C and pressure of 300~Pa for all the different grown films. In each growth run, we also use a Si wafer without SiO\textsubscript{2} as a reference, as this simpler layer structure limits the number of free parameters in fitting of ellipsometer measurements of the SiN thickness and refractive index. The ellipsometer measurements are performed over a wavelength range of 600~nm to 1700~nm, and are carried out for films grown with NH$_3$:DCS ratios of 3:1, 5:1, 7:1 and 15:1 (\cref{fig:2}(a)). We fit the ellipsometer measurements to a Sellmeier model, which has been found to be accurate for the SiN refractive index over a large bandwidth~\cite{sellmeierUeberDurchAetherschwingungen1872} $ \varepsilon_\mathrm{SiN}(\lambda_\mu) = n_\mathrm{SiN}^2(\lambda_\mu) = 1+ A_1 \frac{\lambda_\mu^2}{\lambda_\mu^2 - B_1^2}  - E_1\lambda_\mu^2$, with $\lambda_\mu = 10^6c/f$ being the wavelength in micrometers, and
$c$ and $f$ being the speed of light and the frequency respectively. We note that other models have been used to fit ellipsometer measurements of SiN films taken over more extended infrared wavelengths~\cite{lukeBroadbandMidinfraredFrequency2015}; however, as our concern in this work is the behavior in the near-infrared and telecom bands, the Sellmeier model is adequate.

A clear trend appears where the larger the NH\textsubscript{3}/DCS gas ratio, which corresponds to a lower proportion of silicon in the chemical process, the lower the refractive index, varying from around $n = 1.998$ and $n = 2.012$ to $n=1.967$ and $n = 1.982$ at 193~THz and 283~THz (\textit{i.e.} 1555~nm and 1060~nm) for 3:1 and 15:1 gas ratios, respectively (\cref{fig:2}(b)). Although the material refractive index variation is important for resonator design to accomplish the desired dispersion for microcomb applications, it is also critical that the resonator intrinsic quality factor remain high, that is, that the differences in material growth do not result in added absorption. We fabricated microring resonators with ring radius $RR=23$~{\textmu}m, a SiN thickness of close to $H=440$~nm, and a ring width $RW=1060$~nm, in each of the SiN films, and performed linear transmission measurements of the resulting cavity in the 280~THz band, retrieving both the coupling and intrinsic quality factors~\cite{borselliRayleighScatteringLimit2005}. The extracted average intrinsic quality factors (\cref{fig:2}(c)), which are in the 10$^6$ range, do not show a clear trend with precursor gas ratio. It is likely that the observed differences are due to process variation from run-to-run, rather than any specific attribute of the SiN films.

\begin{figure}[t]
    \begin{center}
        \includegraphics{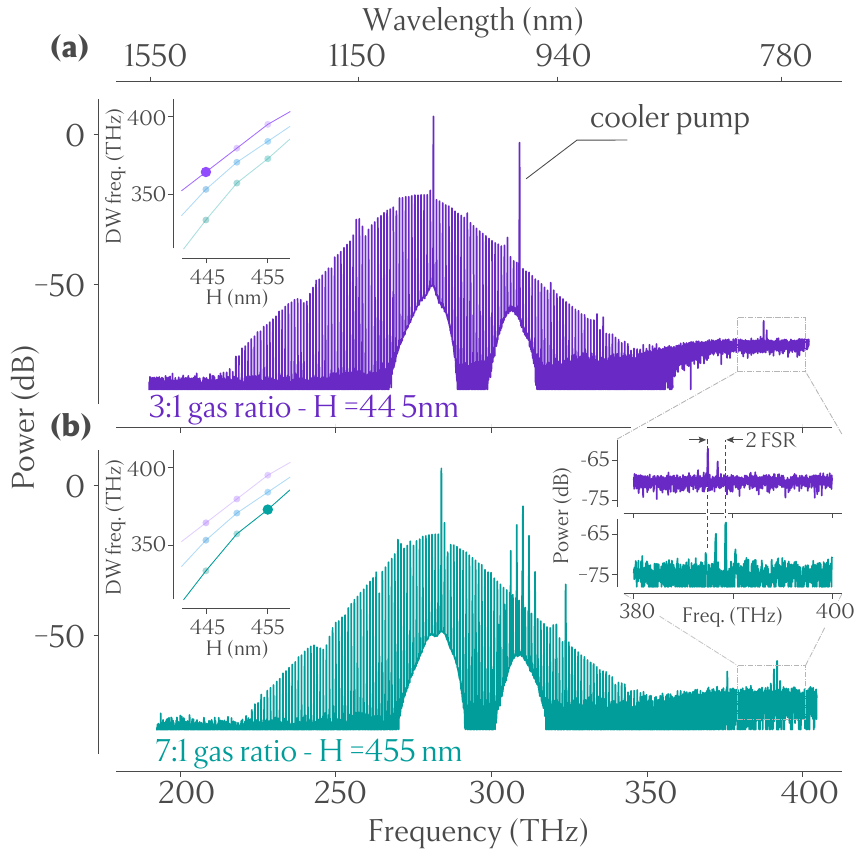}
        \caption{\label{fig:4} DKS generated for the (a) 3:1 and (b) 7:1 gas ratios, exhibiting a short wavelength DW whose frequency matches to within 2~THz (right inset), as expected from the dispersion simulation presented in \cref{fig:3}(c) for each thickness (left insets).}
    \end{center}
\end{figure}

% --- Dispersion engineering of gas ratio -- 
% ------------------------------------------------------------------
The cavity dispersion is composed of contributions from both geometric and material dispersion, and we next consider how it is impacted by the variation in SiN precursor gas ratio. We use finite element method (FEM) simulations to compute the resonator dispersion characteristics for a variety of resonator geometries with different refractive index dispersion profiles based on the data from fig.~\ref{fig:2}. In this study, the microring with a $RR=23$~{\textmu}m ring radius and a $RW=1060$~nm ring width is simulated with the thickness varied from $H=410$~nm to $H=485$~nm for the different gas ratios presented above. Because the refractive index is changed through the precursor gas ratio, for a fixed geometry - and hence a fixed geometric dispersion - the total ring dispersion is changed (\cref{fig:3}(a)). For example, the ring exhibit complete normal dispersion at a thickness below 410~nm for the 3:1 gas ratio and a thickness below 415~nm for the 5:1 gas ratio. The impact of the material dispersion is made more clear by studying the integrated dispersion, defined as the discrepancy between the cavity resonances and a fixed frequency grid defined by the free spectral range (FSR) around the pump, that is, $D_\mathrm{int}(\mu)=\omega_\mathrm{res} - (\omega_\mathrm{pmp} + D_\mathrm{1}\mu)$, with  $\omega_\mathrm{res,pmp}$ being the angular frequency of resonance and the pumped one respectively, $D_\mathrm{1} = 2\pi \times \mathrm{FSR}$, and $\mu$ is the mode number relative to the pumped mode (\textit{i.e} $\omega_\mathrm{res}(\mu=0) = \omega_\mathrm{pmp}$). Studying the integrated dispersion (\cref{fig:3}(b)) allows for a more intuitive prediction of the supported DKS behavior, as $D_\mathrm{int}=0$ correspond to the location of DKS-induced dispersive waves (DWs), whose existence helps to broaden the comb spectrum beyond the anomalous dispersion window~\cite{gaeta_photonic-chip-based_2019}, and which can be tailored to aid in comb self-referencing~\cite{liOctavespanningMicrocavityKerr2015}. From this standpoint, the impact of the material dispersion is obvious as it significantly modifies the frequency position of the DWs,  exhibited through the integrated dispersion zero crossings and the resulting predicted comb spectra based on Lugiato-Lefever Eqaution (LLE) modeling~\cite{moillePyLLEFastUser2019}. For example, at the same thickness $H=455$~nm, the high frequency DW exhibits a shift of more than 10~THz from 3:1 to 7:1 gas ratio, ultimately decreasing the bandwidth from an octave spanning microcomb with a span of 227~THz to a span of close to 100~THz. Interestingly, it is possible to at least partially compensate for changes in the material dispersion that arise from different growth conditions through appropriate adjustment of the geometric dispersion. For instance, it is possible to compensate for variations in the spectral position of one DW due to different gas ratios (\cref{fig:3}(c)) by adjusting the ring thickness. Nearly the same high frequency (short wavelength) DW can be obtained for any of the 3:1, 5:1, or 7:1 gas ratios, by adjusting the thickness between 445~nm and 453~nm. However, it is important to note that the thickness alone cannot compensate for shifts in the position of both DWs, and adjustments in the ring width would also be needed to provide the two-dimensional parameter space needed for  such compensation.

To demonstrate such compensation of changes in material dispersion with adjustments to the geometric dispersion, we study two microring resonators with nominally identical in-plane dimensions, but different gas ratios were used in the SiN film growth (3:1 and 7:1), along with a thickness difference of about 10~nm (445 and 455~nm, respectively). We pump both resonators around the same frequency of 281~THz (i.e. 1067~nm), corresponding to the same fist order transverse electric (TE\textsubscript{0}) mode, with an in-waveguide power of $P_\mathrm{pmp} = 120$~mW. In order to reach the single soliton state, we actively cool the resonator by counterclockwise pumping a cross-polarized mode at a different frequency (307~THz or 977~nm), which has been demonstrated to allow adiabatic access to DKS states~\cite{zhouSolitonBurstsDeterministic2019,moilleUltraBroadbandSolitonMicrocomb2021}. The obtained DKS spectrum for each ring resonator exhibits almost the same high frequency DW (\cref{fig:4}(a)-(b)), to within 2~THz (i.e., a discrepancy of only 2 comb teeth), as predicted from LLE simulations where only the thickness and the material dispersion have been changed. In contrast, if the gas ratio was kept fixed (e.g., at 7:1), a 10~nm thickness change, from 445~nm to 455~nm, is predicted to result in a shift of the DW from approximately 335~THz to 365~THz corresponding to a $\approx$30~FSRs shift of the short wavelength DW. 

In conclusion, we have demonstrated that even when operating entirely within a regime of LPCVD growth in which silicon nitride films are commonly considered to be stoichiometric, changes to the gas ratio continue to impact chromatic dispersion and its influence on microresonaor frequency comb generation. Although the change in material dispersion is not as dramatic as when moving to the silicon-rich regime, it is substantial enough to significantly impact the properties of broadband microresonator frequency. Along with studying the impact of precursor gas ratio on material dispersion and integrated dispersion of microring resonators fabricated from these films, we show that changes in geometric dispersion, for example resonator thickness, can be used to compensate for changes in material dispersion. Going forward, we believe our results point to the ability to use material growth conditions together with geometric control in the dispersion engineering of microresonators for broadband frequency comb applications.

\noindent \textbf{Funding.}~Defense Advanced Research Projects Agency (DARPA-APHI); National Institute of Standards and Technology (NIST-on-a-chip). This work was performed in part at the Cornell NanoScale Facility, which is supported by the National Science Foundation (Grant NNCI-2025233).

\noindent\textbf{Disclosures.}~The authors declare no conflicts of interest.
{\bibliographystyle{osajnlnt}
\bibliography{Biblio}
\ifthenelse{\equal{\journalref}{ol}}{%
\clearpage
\bibliographyfullrefs{BiblioFull}
}{}
}

\end{document}